\documentclass[onecolumn,showpacs,prd,floatfix,axodraw]{revtex4}
\usepackage{mathrsfs}
\usepackage{graphicx,,booktabs,bm}
\usepackage{overpic}
\usepackage{color}
\usepackage{amssymb}

\usepackage{mathrsfs,bm,amsmath,amssymb}
\usepackage{longtable,lscape}
\usepackage{txfonts}
\usepackage{amssymb}
\usepackage{indentfirst}
\usepackage{graphicx,,booktabs}
\usepackage{multirow,ulem}

\usepackage{color}
\usepackage{amssymb}

\definecolor{cover}{rgb}{0.77,0.87,0.88}
\definecolor{blueone}{rgb}{0.1,0.1,.7}
\definecolor{citec}{rgb}{0.14,0.47,0.09}
\definecolor{two}{rgb}{0.0,0.5,0.}
\definecolor{three}{rgb}{.5,.1,0.15}
\usepackage[bookmarks=true,bookmarksopen=false,plainpages=false,breaklinks=true,
   bookmarksnumbered=true,hypertexnames=false,
   filecolor=blue,urlcolor=three,menucolor=three,
   linkcolor=three,citecolor=blueone, colorlinks,
   anchorcolor=blue,runcolor=pink,frenchlinks=red
   pdfstartview=FitH,pdftitle=title,%
   pdfauthor=author]{hyperref}

\begin{document}
\title{Investigating the Molecular Nature of $\Lambda_b(6146)$ and $\Lambda_b(6152)$ via Strong Decays and Predicting Their Heavy-Quark Symmetry Partners}
\author{Jing-wen Feng}
\affiliation{School of Physics and Electronic Engineering, Sichuan Normal University, Chengdu 610101, China}
\author{Cai Cheng\footnote{corresponding author}} \email{ccheng@sicnu.edu.cn}
\affiliation{School of Physics and Electronic Engineering, Sichuan Normal University, Chengdu 610101, China}
\author{Yin Huang\footnote{corresponding author}} \email{huangy2019@swjtu.edu.cn}
\affiliation{School of Physical Science and Technology, Southwest Jiaotong University, Chengdu 610031,China}

\begin{abstract}
Heavy quark symmetry can help us identify the internal structure of hadrons and predict new particles. In this study, we examine the strong decay modes of the observed $\Lambda_b^0(6146)$ and $\Lambda_b^0(6152)$, assuming these two states are molecular states primarily composed of $\bar{B}^{*}N$ component. The partial decay widths of the $\bar{B}^{*}N$ molecular state into the $\pi\Sigma_b$ and $\pi\Sigma_b^{*}$ final states through hadronic loops are calculated using effective Lagrangians.  Our results, when compared with LHCb observations, support the interpretation of $\Lambda_b^0(6146)$ as a molecule primarily composed of $\bar{B}^{*}N$ components. However, the decay width of $\Lambda_b^0(6152)$ cannot be accurately reproduced within the molecular state framework.  Based on the above results and heavy quark symmetry, we predict the existence of $\bar{B}^{*}N$ molecular states with $J^p=5/2^{+}$, which are the heavy quark spin symmetry partners of $\Lambda_b(6146)$, with masses in the range of 6195-6200 MeV.  And the main decay is $\pi\Sigma_b^{*}$ channel.  Moreover, there must existence of a $D^{*}N$ molecule with $J^p=3/2^{+}$, possible corresponding to the experimentally observed $\Lambda_c(2860)^{+}$. If $\Lambda_c(2880)^{+}$ is indeed the heavy quark flavor symmetry partner of $\Lambda_b(6152)$, it would exhibit a conventional three-quark structure. Therefore, we also propose the search for a $D^{*}N$ molecule with a spin-parity of $J^p=5/2^{+}$, which would be the heavy-quark spin partner state of $\Lambda_c(2860)^{+}$. It should be noted that these baryons may be mixed states, containing both molecular and three-quark components.  These results can aid experiments in exploring the internal structure of these baryons.
\end{abstract}

\date{\today}

%\pacs{{13.60.Le}{inner structures}   \and {13.85.Lg}{Schr\"{o}dinger equation} \and  {25.30.-c}{molecular state}  }}

\maketitle
\section{Introduction}\label{sec:intro}
In 1964, Zweig made the pioneering prediction of pentaquark states~\cite{Zweig:1964ruk}; however, the confirmation of the existence of such hadronic structures, specifically
$P_c(4450)$ and $P_c(4380)$, did not occur until 2015. This observation was reported by the LHCb Collaboration in the $J/\psi{}p$ final state from the $\Lambda_{b}^{0}\rightarrow J/\psi K^{-}p$
reaction~\cite{LHCb:2015yax}. From the observed $J/\psi{}p$ decay mode, the quark composition of these two states is $uudc\bar{c}$, which is referred to as hidden-charm pentaquark
states.  Due to the limitations of the experimental data accumulated at the time, LHCb Collaboration was unable to fully confirm the spin-parity of these two particles. Therefore,
in 2019, four years later, the LHCb Collaboration analyzed more data collected from the $\Lambda_{b}^{0}\rightarrow J/\psi K^{-}p$ reaction~\cite{LHCb:2019kea}.  The results revealed
the existence of a new hidden-charm pentaquark state, $P_c(4312)$, and identified $P_c(4450)$ as an overlap of two other pentaquark states, $P_c(4440)$ and $P_c(4450)$.  Unfortunately,
despite analyzing a larger dataset than that accumulated by the LHCb Collaboration in 2015, the spin-parity of these pentaquark states has still not been fully determined.

Therefore, people turn their attention to various theoretical models, hoping to compare calculated results such as mass spectra and decay properties with experimental data to determine
their spin-parity. However, this is not an easy task, as these different theoretical models can all provide reasonable explanations for hadrons, even if they all predict a specific
spin-parity for the given hadron.  Here, we take the example of two bottom baryons, $\Lambda_b(6146)$ and $\Lambda_b(6152)$, which we are studying in this paper.  They were first observed
by the LHCb Collaboration in the $\Lambda_b\pi^{+}\pi^{-}$ invariant mass spectrum in 2019~\cite{LHCb:2019soc}.  The observed resonance masses, widths, isospin ($I$), and spin-parity ($J^P$)
are, respectively,
\begin{align*}
 \mathrm{M}_{\Lambda_{b}(6146)}&=6146.17\pm0.33\pm0.22\pm0.16 \ \mathrm{MeV}\\
\mathrm{M}_{\Lambda_{b}(6152)}&=6152.51\pm0.26\pm0.22\pm0.16 \ \mathrm{MeV}\\
\Gamma_{\Lambda_{b}(6146)}&=2.9\pm1.3\pm0.3
 \ \mathrm{MeV} , I(J^{p})=0~(3/2+)\\
\Gamma_{\Lambda_{b}(6152)}&=2.1\pm0.8\pm0.3
 \ \mathrm{MeV} , I(J^{p})=0~(5/2+).
\end{align*}

Considering their masses and small splitting, the LHCb Collaboration suggested that both of them may belong to the doublet of $\Lambda_b(1D)$ states in the quark model. Indeed, by studying
their strong decays, the observed $\Lambda_b(6146)$ and $\Lambda_b(6152)$ might be explained as the $\lambda$-mode $\Lambda_b(1D)$ states in the quark model~\cite{Wang:2019uaj}. In Ref.~\cite{Azizi:2020tgh},
the mass and two-body strong decays of the $\Lambda_b(6146)$ were investigated using the QCD sum rules approach. It was shown that the $\Lambda_b(6146)$ can be considered as $1D$-wave baryons
with $J^p=3/2^{+}$.  In the heavy quark light diquark model, based on the analysis of the two-body Okubo-Zweig-Iizuka (OZI) allowed strong decays, the $\Lambda_b(6146)$ and $\Lambda_b(6152)$
were suggested to be $1D$ bottom baryon excited states with spin-parity $J^p=3/2^{+}$ and $J^p=5/2^{-}$~\cite{Chen:2019ywy}, respectively.  Other explanations for $\Lambda_b(6146)$ and $\Lambda_b(6152)$
as conventional baryons containing a $b$ quark can be found in other works, such as in Refs.~\cite{Liang:2019aag,Yu:2021zvl,Kakadiya:2021jtv}.

Although the studies of Refs.~\cite{Wang:2019uaj,Azizi:2020tgh,Chen:2019ywy,Liang:2019aag,Yu:2021zvl,Kakadiya:2021jtv} seem to indicate that $\Lambda_b(6146)$ and $\Lambda_b(6152)$  are conventional
three-quark state, these two states might still be hadronic molecule state.   One reason is that the mass gap between $\Lambda_b(6146/6152)$ and the ground state $\Lambda_b$, which is about 530 MeV,
is large enough to allow for the excitation of a light quark-antiquark pair, forming a molecular state. If a $u\bar{u}$ quark pair is produced, these two states may be considered as molecules containing
$\bar{B}^{(*)}N$ components. Indeed, it has been demonstrated in Ref.~\cite{Jian:2022rln} that the interaction between a $\bar{B}^{(*)}$ meson and an $N$ baryon is sufficiently strong to form several
bound states, two of which can be identified as the experimentally observed $\Lambda_b(6146)$ and $\Lambda_b(6152)$.  Another reason is that we found the decay widths calculated based on the three-quark
structure are larger than the experimental measurements~\cite{Wang:2019uaj,Chen:2019ywy,Liang:2019aag} (see Tab.~\ref{tab-2}), suggesting that $\Lambda_b(6146)$ and $\Lambda_b(6152)$ might be more
complex structures than quark states, possibly containing other components, such as $\bar{B}^{(*)}N$ molecular components.  What convinces us that the experimentally observed $\Lambda_b(6146)$ and
$\Lambda_b(6152)$ might contain molecular components is the calculation by the authors in Ref.~\cite{Yao:2018jmc}, which found the masses of $\lambda$-mode $D$-wave states with  spin-parities $J^p=3/2^{+}$
and $J^p=5/2^{+}$ to be $\Lambda_b(6190)$ and $\Lambda_b(6196)$, respectively, rather than the experimentally reported $\Lambda_b(6146)$ and $\Lambda_b(6152)$.

\begin{table}[h!]
\centering
\caption{The experimentally observed $\Lambda_c^{+}$ and $\Lambda_b^0$ baryons, as well as possible molecular state interpretations.
The $?$ indicates that there is currently no research regarding considering them as corresponding to molecular state structures.}\label{tab-1}
\setlength{\tabcolsep}{4mm}{
\begin{tabular}{ccc|cccc}
\hline\hline
~~~~~ State                     &~~~~~ $J^p$       &~~~~~ Molecular           &~~~~~State                & ~~~~~$J^p$     & Molecular \\ \hline
~~~~~$\Lambda_c(2595)^{+}$      &~~~~~ $1/2^{-}$   &~~~~~ $DN~$\cite{Nieves:2019nol,Garcia-Recio:2008rjt,Nieves:2019kdh,Guo:2019ytq,Kawakami:2018olq,Lu:2014ina,Tolos:2004yg,Hofmann:2005sw,Mizutani:2006vq,Jimenez-Tejero:2009cyn}        &~~~~~$\Lambda_b(5912)^0$  &~~~~~ $1/2^{-}$   &~~~~~ $\bar{B}N$\cite{Kawakami:2018olq,Lu:2014ina,Garcia-Recio:2012lts,Liang:2014eba,Huang:2021ave}          \\
~~~~~$\Lambda_c(2625)^{+}$      &~~~~~ $3/2^{-}$   &~~~~~ $D^{*}N~$\cite{Nieves:2019nol,Garcia-Recio:2008rjt,Nieves:2019kdh,Guo:2019ytq,Kawakami:2018olq,Lu:2014ina}      &~~~~~$\Lambda_b(5920)^0$  &~~~~~~~$3/2^{-}$ & $~~~~~\bar{B}^{*}N$\cite{Kawakami:2018olq,Lu:2014ina,Garcia-Recio:2012lts,Liang:2014eba,Huang:2021ave}         \\
~~~~~$\Lambda_c(2860)^{+}$      &~~~~~ $3/2^{+}$   &~~~~~ $D^{(*)}N~$? &~~~~~$\Lambda_b(6146)^0$ &~~~~~$3/2^{+}$  &~~~~~$\bar{B}^{(*)}N$~\cite{Jian:2022rln}\\
~~~~~$\Lambda_c(2880)^{+}$      &~~~~~ $5/2^{+}$   &~~~~~ $D^{(*)}N~$? &~~~~~$\Lambda_b(6152)^0$ &~~~~~$5/2^{+}$  &~~~~~$\bar{B}^{(*)}N$~\cite{Jian:2022rln}\\
\hline \hline
\end{tabular}}
\end{table}
A more significant reason is based on heavy quark symmetry~\cite{Isgur:1991wq}, which incorporates both heavy quark spin symmetry and heavy quark flavor symmetry.  Experimentally, several charm-containing
baryons $\Lambda_c^{+}$ have been observed~\cite{ParticleDataGroup:2022pth}.   Among them, $\Lambda_c(2595)^{+}$ and $\Lambda_c(2625)^{+}$ can be interpreted as molecular states mainly composed of $\bar{D}N$
and $\bar{D}^{*}N$ components~\cite{Nieves:2019nol,Garcia-Recio:2008rjt,Nieves:2019kdh,Guo:2019ytq,Kawakami:2018olq,Lu:2014ina,Tolos:2004yg,Hofmann:2005sw,Mizutani:2006vq,Jimenez-Tejero:2009cyn}, respectively
(see Tab.~\ref{tab-1} ), forming heavy quark spin symmetry (HQSS) doublet states.  Simultaneously, heavy quark spin symmetry also implies the existence of other high spin-parity HQSS doublet states corresponding
to the $D^{(*)}N$ system, with their spin-parity being $J^p=3/2^{+}$ and $J^{p}=5/2^{+}$, respectively~\cite{Yamaguchi:2014era}. After carefully examining the PDG~\cite{ParticleDataGroup:2022pth}, it is
found that these two molecular states may correspond to $\Lambda_c(2860)^{+}$ and $\Lambda_c(2880)^{+}$, respectively.  However, there is currently no research on whether they can be considered hadrons containing
molecular components.  According to heavy quark flavor symmetry, we propose a scheme to investigate whether they can be considered as $D^{(*)}N$ molecules, which involves studying their heavy quark flavor
partners.  Thus, we hope that $\bar{B}^{(*)}N$ molecules exist.

Fortunately, the experimentally observed $\Lambda_b(5912)^0$ and $\Lambda_b(5920)^0$ can be regarded as molecules
containing $\bar{B}N$ and $\bar{B}^{*}N$ components~\cite{Kawakami:2018olq,Lu:2014ina,Garcia-Recio:2012lts,Liang:2014eba,Huang:2021ave}, respectively. They can be respectively accepted as the heavy-quark flavor partners for $\Lambda_c(2595)^{+}$ and $\Lambda_c(2625)^{+}$. The $\Lambda_b(6146)^0$ and $\Lambda_b(6152)^0$, which could be accepted as $\bar{B}^{(*)}N$ molecules~\cite{Jian:2022rln}, might respectively be the heavy-quark flavor partner states of $\Lambda_c(2860)^0$ and $\Lambda_c(2880)^0$ due to the heavy-quark flavor partners are expected to have the same spin-parity(see Tab.~\ref{tab-1}).  Therefore, confirming whether $\Lambda_b(6146)^0$ and $\Lambda_b(6152)^0$ can be interpreted as molecular states is crucial for our understanding of the molecular structures of $\Lambda_c(2860)^0$ and $\Lambda_c(2880)^0$.

\begin{figure}[h!]
\centering
	\includegraphics[width=0.55\linewidth]{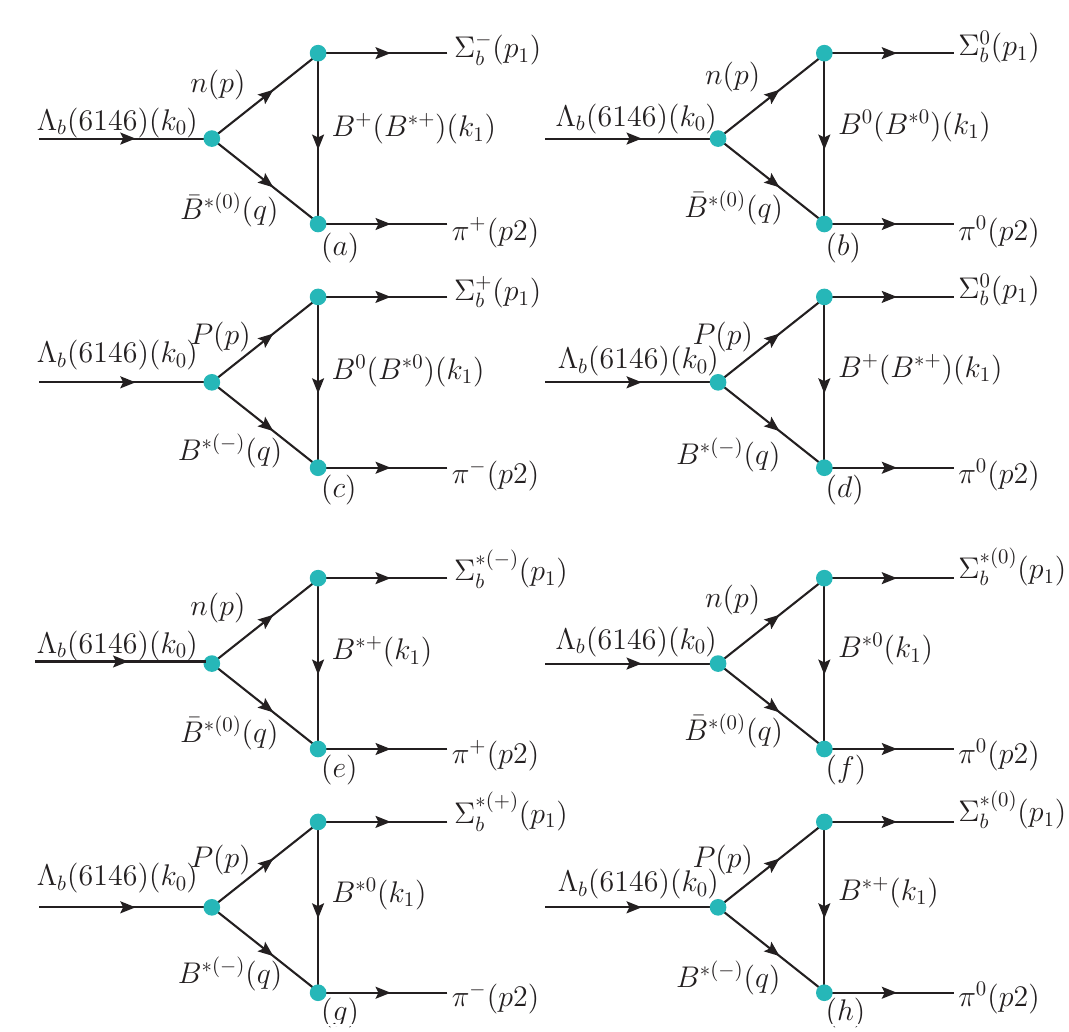}
	\caption{Feynman diagram for the $\Lambda_{b}(6146) $$\to$$\Sigma_{b}^{(*)}\pi$  decay processes. The contributions from the t-channel $B$ and $B^{*}$ exchange are considered.We also show the definition of the kinematical $(k0, p, q, p_{1}, p_{2}, k_{1})$ that we use in the present calculation.}\label{a1}
\end{figure}
	\begin{figure}[h!]
\centering
	\includegraphics[width=0.55\linewidth]{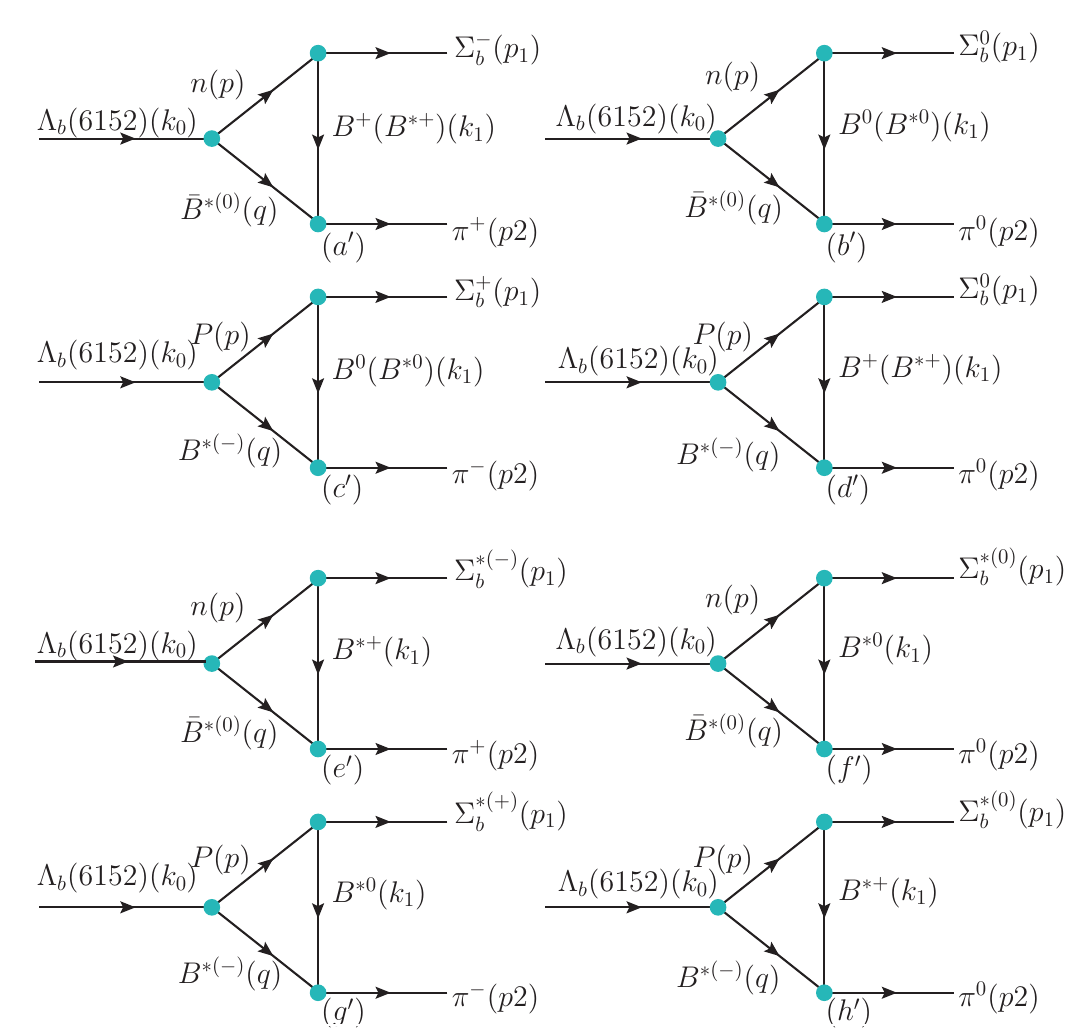}
	\caption{Feynman diagram for the $\Lambda_{b}(6152) $$\to$$\Sigma_{b}^{(*)}\pi$ decay processes.}
	\label{a2}
\end{figure}

At present, the $\bar{B}^{(*)}N$ molecular interpretation of $\Lambda_b(6146)^0$ and $\Lambda_b(6152)^0$ is primarily supported by examinations of their mass spectra. However, extensive research, such as in Refs.~\cite{Zhu:2021exs,Huang:2018wgr}, indicates that relying solely on mass spectra studies may not conclusively confirm their molecular state structure. Further examination of their decay widths is also required.
In this work, based on the findings from Ref.~\cite{Jian:2022rln}, we propose that $\Lambda_b(6146)^0$ and $\Lambda_b(6152)^0$ are $\bar{B}^{(*)}N$ molecular states, and investigate their strong decay widths to confirm this interpretation. Next, we will introduce the theoretical framework and conclude with numerical results and discussions.

\section{FORMALISM AND INGREDIENTS}\label{Sec: formulism}
In this work, we calculate the strong decay widths of the $\Lambda_{b}(6146)$ and $\Lambda_{b}(6152)$ by considering them as $\bar{B}^{(*)}N$ molecules proposed by Ref.~\cite{Jian:2022rln}.
These two states were observed by the LHCb Collaboration in the $\Lambda_b\pi^{+}\pi^{-}$ invariant mass spectrum through intermediate states $\Sigma_b^{(*)}\pi$\cite{LHCb:2019soc}. Hence,
we only compute their $\Sigma^{(*)}_b\pi$ decay widths. Since the baryons $\Sigma^{(*)}_b$ almost exclusively decay to the $\Lambda_b\pi$ final state\cite{ParticleDataGroup:2022pth}, we can
directly compare the computed results with experimental data to assess their molecular state structure. From Ref.~\cite{Jian:2022rln}, it can be observed that the $\bar{B}^{*}N$ molecular
component plays a predominant role in these two states, accounting for approximately 93\%-99\% of the total component. Therefore, we only calculate their decay widths through the $\bar{B}^{*}N$
molecular component decaying into $\pi\Sigma_b^{(*)}$ final states. The relevant Feynman diagrams for the processes are shown in Fig. \ref{a1} and Fig. \ref{a2}. It is found that the contribution
mainly comes from the $t$-channel $B$ and $B^{}$ mesons exchange.

To compute the decay widths shown in Fig.\ref{a1} and Fig.\ref{a2}, we need the effective Lagrangian densities for relevant interaction vertices.  Due to the spin-parity of $\Lambda_b(6146)$ and
$\Lambda_b(6152)$ being $J^P=3/2^{+}$ and $J^P=5/2^{+}$ respectively, their coupling interactions with the $\bar{B}^{*}N$ component should predominantly occur via the $P$-wave and $F$-wave,
respectively.  Their simplest forms can be found in Ref.~\cite{Dong:2016dkh}, and have following form,
\begin{align}
\mathcal{L}_{\Lambda_{b}^{*}N\bar{B}^{*}}^{3/2^{+}}(x)&=-i \sum_{j=n\bar{B}^{*0},pB^{*-}}\xi_{j}g_{\Lambda_{b}^{*}N\bar{B}^{*}} \int d^{4}y\Phi(y^2)N(x+\omega_{\bar{B}^{*}}y)\gamma^5\bar{B}^{*}_{\mu}(x-\omega_{N}y)\Lambda_{b}^{*\mu}(x),\label{eqa11}\\
 \mathcal{L}_{\Lambda_{b}^{*}N\bar{B}^{*}}^{5/2^{+}}(x)&=\sum_{j=n\bar{B}^{*0},pB^{*-}}\xi_{j}g_{\Lambda_{b}^{*}N\bar{B}^{*}}\int d^{4}y\Phi(y^2)N(x+\omega_{\bar{B}^{*}}y)\partial _{\mu}\bar{B}^{*}_{\nu}(x-\omega_{N}y)\Lambda_{b}^{*\mu\nu}(x) \label{eq1},
\end{align}
where $\omega_{\bar{B}^{*}}=m_{\bar{B}^{*}}/(m_{\bar{B}^{*}}+m_{N})$ and $\omega_{m_{N}}=m_{N}/(m_{\bar{B}^{*}}+m_{N})$ with $m_{\bar{B}^{*}}$  and $m_{N}$ are the masses of the meson $\bar{B}^{*}$
and baryon $N$, respectively.  The values of $\xi_{j}$, defined as
\begin{equation}
\xi_{j}=
\begin{cases}
-1/\sqrt{2} & j=n\bar{B}^{*0} \\
	 \;\;1/\sqrt{2} & j=pB^{*-}\label{eq3},
\end{cases}
\end{equation}
are determined based on isospin assignments.  In the Lagrangians, an effective correlation function $\Phi(y^{2})$ is introduced to depict the spatial distribution of the two constituents,
namely, the $\bar{B}^{*}$ and the $N$, within the hadronic molecular states $\Lambda_b(6146)$ and $\Lambda_b(6152)$. This correlation function is also instrumental in rendering the Feynman
diagrams ultraviolet finite. Specifically, we adopt the Fourier transform of the correlation function to follow a Gaussian distribution in Euclidean space~\cite{Huang:2018wgr,Zhu:2021exs,Xiao:2016mho,Dong:2016dkh,Dong:2008gb,Dong:2017rmg}:
\begin{align}
    \Phi(p_{E}^{2}/\Lambda^{2})\doteq \textup{exp}\, (-p_{E}^{2}/\Lambda^{2})\label{eq4},
\end{align}
where $p_{E}$ denotes the Euclidean Jacobi momentum, and $\Lambda$ represents the size parameter that characterizes the distribution of the components within the molecule.  The precise value of $\Lambda$ cannot be determined from first principles; therefore, it is typically derived from experimental data. It is commonly chosen to be around 1 GeV, primarily because this choice often leads to computed results that are consistent with experimental observations~\cite{Huang:2018wgr,Zhu:2021exs,Xiao:2016mho,Dong:2016dkh,Dong:2008gb,Dong:2017rmg}. In this study, we vary $\Lambda$ within the range of 0.9 GeV to 1.10 GeV to investigate whether $\Lambda_b(6146)$ and $\Lambda_b(6152)$ can be interpreted as $\bar{B}^{*}N$ molecules.

In the above equations, the coupling constants $g_{\Lambda_b^{*}N\bar{B}^{*}}$ with different spin-parity assignments are unknown and can be computed using the compositeness condition \cite{Weinberg:1962hj,Salam:1962ap}.
This condition implies that the renormalization constants of a molecular wave function should be zero, i.e.,
\begin{align}
    Z_{\Lambda_{b}^{*}}&=1-\frac{d\Sigma_{\Lambda_{b}^{*}}^{T}(k_{0})}{dk\!\!\!/_0}|_{k\!\!\!/_0=m_{\Lambda_{b}^{*}}}=0 \qquad &j&=\frac{3}{2},\frac{5}{2}\label{eq8},
\end{align}
where the $\Sigma^{T}_{\Lambda_{b}^{*}}(k_{0})$ denotes the transverse part of the mass operator, which is related to its mass operator by the relation:
\begin{align}
    \Sigma_{\Lambda_{b}^{*}}^{\mu\nu}(k_{0})&=(g_{\mu\nu}-\frac{k_{0}^{\mu}k_{0}^{\nu}}{k_{0}^{2}})\Sigma_{\Lambda_{b}^{*}}^{T}(k_{0})+\cdots
    &j=\frac{3}{2},\\
    \Sigma_{\Lambda_{b}^{*}}^{\mu\nu\alpha\beta}(k_{0})&=\frac{1}{2}(g^{\mu\alpha}_{\bot}g^{\nu\beta}_{\bot}+g^{\mu\beta}_{\bot}g^{\nu\alpha}_{\bot})\Sigma^{T}_{\Lambda_{b}^{*}}(k_{0})+\cdots\
    &j=\frac{5}{2}.
\end{align}
\begin{figure}[h!]
\center
	\includegraphics[width=0.35\linewidth]{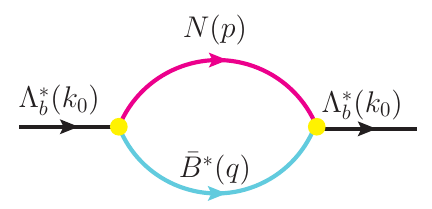}
\center	
\caption{Slef-energy of the $\Lambda_{b}^{*}$.}
	\label{a3}
\end{figure}
The concrete form of the mass operators of the $\Lambda_b^{*}$ corresponding to the diagrams in Fig.~\ref{a3} are
\begin{align}
\Sigma^{3/2+}_{\Lambda_{b}(6146)}&=\sum_{j}\xi_{j}(g_{\Lambda_{b}^{*}N\bar{B}^{*}}^{3/2+})^{2}\int\frac{d^4q}{(2\pi)^4}\Phi^2[(p\omega_{\bar{B}^{*0}}-q\omega_{n})]\gamma_{\mu}\gamma_5\frac{k\!\!\!/_0-q\!\!\!/+m_{N}}{(k_0-q)^2-m_N^2}\gamma_5\gamma_{\nu}\frac{-g^{\mu\nu}+q^{\mu}q^{\nu}/m^2_{\bar{B}^{*}}}{q^2-m^2_{\bar{B}^{*}}}\nonumber\\
                                 &=(g_{\Lambda_{b}^{*}N\bar{B}^{*}}^{3/2+})^{2}\int_{0}^{\infty} d\alpha \int_{0}^{\infty}d\beta \sum_{j}\xi_{j}\mathcal{W}(\omega_{\bar{B}^{*}},m_{\bar{B}^{*}})[\frac{\Lambda^{2}\mathcal{Q}(\omega_{\bar{B}^{*}})}{2\mathcal{P}(m_{\bar{B}^{*}})z}+\frac{k_{0}\Lambda^{2}}{\mathcal{P}(m_{\bar{B}^{*}})}+\frac{\mathcal{Q}(\omega_{\bar{B}^{*}})}{2z}+k_{0}\nonumber\\
                                 &-\frac{\Lambda^{2}m_{N}}{\mathcal{P}(m_{\bar{B}^{*}})}-m_{N}](g_{\mu\nu}-\frac{k_{0}^{\mu}k_{0}^{\nu}}{k_{0}^{2}})+....\label{eq11},\\
\Sigma^{\mu\nu\alpha\beta}_{\Lambda_{b}(6152)}&=\sum_{j}\xi_{j}(g_{\Lambda_{b}^{*}N\bar{B}^{*}}^{5/2+})^{2}\int\frac{d^4q}{(2\pi)^4}\Phi^2[(p\omega_{\bar{B}^{*0}}-q\omega_{n})]\frac{k\!\!\!/_0-q\!\!\!/+m_{N}}{(k_0-q)^2-m_N^2}\frac{-g^{\mu\alpha}+q^{\mu}q^{\alpha}/m^2_{\bar{B}^{*}}}{q^2-m^2_{\bar{B}^{*}}}q^{\nu}q^{\alpha}\nonumber
                                  \end{align}
\begin{align}
                                  &=(g_{\Lambda_{b}^{*}N\bar{B}^{*}}^{5/2+})^{2}\int_{0}^{\infty} d\alpha \int_{0}^{\infty}d\beta \sum_{j}\xi_{j}\mathcal{W}(\omega_{\bar{B}^{*}},m_{\bar{B}^{*}})[\frac{\mathcal{Q}(\omega_{\bar{B}^{*}})\Lambda^{4}}{8\mathcal{P}(m_{\bar{B}^{*}})z^{2}}+\frac{k_{0}\Lambda^{4}}{4\mathcal{P}(m_{\bar{B}^{*}})z}+\frac{\Lambda^{4}m_{N}}{4\mathcal{P}(m_{\bar{B}^{*}})z}]\frac{1}{2}(g^{\mu\alpha}_{\bot}g^{\nu\beta}_{\bot}+g^{\mu\beta}_{\bot}g^{\nu\alpha}_{\bot})+....,\label{eq10}
\end{align}

with
\begin{align}
\mathcal{Q}(\omega_{\bar{B}^{*}})=&(-4\omega_{\bar{B}^{*}}-2\beta)k_{0},\mathcal{P}(m_{\bar{B}^{*}})=2m^{2}_{\bar{B}^{*}}z,\nonumber\\
\mathcal{W}(\omega_{\bar{B}^{*}},m_{\bar{B}^{*}})=&\frac{1}{16\pi^{2}z^{2}}{\rm exp}\{-\frac{1}{\Lambda^{2}}[-2k_{0}^{2}\omega_{\bar{B}^{*}}^{2}
+\alpha m_{\bar{B}^{*}}^{2}+\beta(-k_{0}^{2}+m^{2}_{N})+\frac{(-4\omega_{\bar{B}^{*}}-2\beta)^{2}k_{0}^{2}}{4z}]\}.\nonumber
\end{align}
where $z=2+\alpha+\beta$ and $k_0^2=m^2_{\Lambda_b^{*}}$ with $m_{\Lambda_b^{*}}$ is the mass of the $\Lambda_b^{*}$.  $m_{\bar{B}^{*}}$ and $m_N$ are the masses of the $\bar{B}^{*}$ and $N$, respectively.

To evaluate the diagrams, in addition to the Lagrangians in Eqs.~(\ref{eqa11},\ref{eq1}), the following effective Lagrangians, responsible for $B^{*}\bar{B}^{*}\pi, B\bar{B}^{*}\pi$, and $\Sigma^{(*)}_{b}N\bar{B}^{(*)}$ interaction vertexes are needed as well.~\cite{Jian:2022rln,Casalbuoni:1996pg,Wise:1992hn,Yan:1992gz,Cheng:1992xi,Xiao:2017uve,Dong:2012hc,Navarra:2001ju}
\begin{align}
\mathcal{L}_{\Sigma_{b}N\bar{B}}&=g_{\Sigma_{b}N\bar{B}}\Sigma_{b}N\bar{B},\nonumber\\
\mathcal{L}_{\Sigma_{b}N\bar{B}^{*}}&=g_{\Sigma_{b}N\bar{B}^{*}}\Sigma_{b}\gamma^{\mu}N\bar{B}^{*}_{\mu},\\
\mathcal{L}_{\Sigma_{b}^{*}N\bar{B}^{*}}&=g_{\Sigma_{b}^{*}N\bar{B}^{*}}\Sigma_{b}^{*\mu}N\bar{B}^{*}_{\mu},\nonumber\\
\mathcal{L}_{B\bar{B}^{*}\pi}&=\frac{2ig\sqrt{m_{B}m_{\bar{B}^{*}}}}{f_{\pi}}B^{-}B^{*0\lambda}\pi^{+(\lambda)}+\frac{\sqrt{2}ig\sqrt{m_{B}m_{\bar{B}^{*}}}}{f_{\pi}}\bar{B}^{0}B^{*0\lambda}\pi^{0(\lambda)}+\frac{2ig\sqrt{m_{B}m_{\bar{B}^{*}}}}{f_{\pi}}\bar{B}^{0}B^{*+\lambda}\pi^{-(\lambda)}+\frac{\sqrt{2}ig\sqrt{m_{B}m_{\bar{B}^{*}}}}{f_{\pi}}B^{+}B^{*-\lambda}\pi^{0(\lambda)},\nonumber\\		
\mathcal{L}_{B^{*}\bar{B}^{*}\pi}&=\frac{g\epsilon_{\alpha\mu\nu\lambda}}{f_{\pi}}B^{*-\lambda}B^{*0(\alpha)\mu}\pi^{+(\nu)}-\frac{g\epsilon_{\alpha\mu\nu\lambda}}{f_{\pi}}B^{*0\mu}B^{*-(\alpha)\lambda}\pi^{+(\nu)}+\frac{g\epsilon_{\alpha\mu\nu\lambda}}{\sqrt{2} f_{\pi}}B^{*0\mu}\bar{B}^{*0(\alpha)\lambda}\pi^{0(\nu)}-\frac{g\epsilon_{\alpha\mu\nu\lambda}}{\sqrt{2} f_{\pi}}\bar{B}^{*0\lambda}B^{*0(\alpha)\mu}\pi^{0(\nu)}\nonumber\\
	&+\frac{g\epsilon_{\alpha\mu\nu\lambda}}{\sqrt{2} f_{\pi}}B^{*0\mu}\bar{B}^{*0(\alpha)\lambda}\pi^{0(\nu)}-\frac{g\epsilon_{\alpha\mu\nu\lambda}}{\sqrt{2} f_{\pi}}\bar{B}^{*0\lambda}B^{*0(\alpha)\mu}\pi^{0(\nu)}+\frac{g\epsilon_{\alpha\mu\nu\lambda}}{\sqrt{2} f_{\pi}}B^{*-\lambda}B^{*+(\alpha)\mu}\pi^{0(\nu)}-\frac{g\epsilon_{\alpha\mu\nu\lambda}}{\sqrt{2} f_{\pi}}B^{*+\mu}B^{*-(\alpha)\lambda}\pi^{0(\nu)}\nonumber\\	
    &+\frac{g\epsilon_{\alpha\mu\nu\lambda}}{f_{\pi}}\bar{B}^{*0\lambda}B^{*+(\alpha)\mu}\pi^{-(\nu)}-\frac{g\epsilon_{\alpha\mu\nu\lambda}}{f_{\pi}}B^{*+\mu}\bar{B}^{*0(\alpha)\lambda
	}\pi^{-(\nu)},\nonumber
\end{align}
where the coupling constants $g_{\Sigma_{b}N\bar{B}}=5.7$, $g_{\Sigma_{b}N\bar{B}^{*}}=3.7$, and $g_{\Sigma_{b}^{*}N\bar{B}^{*}}=6.9$ are obtained from Ref.\cite{Torres-Rincon:2014ffa}. The parameters involved here are $f_{\pi}=132$ MeV and $g=0.59$, determined from Refs.\cite{Casalbuoni:1996pg,Yue:2022gym}.

By putting all the pieces together, the amplitudes corresponding to the Feynman diagrams in Figs.~\ref{a1} and \ref{a2} can be readily computed and are written as follows
\subsection{The decay $\Lambda_{b}(6146)\to{}\Sigma_{b}^{(*)}\pi$}
\begin{align}
\mathcal{M}_{a}^{3/2+}&=\frac{1}{\sqrt{2}}i\mu(p_{1})\{ig_{\Lambda_{b}^{*} n\bar{B}^{*0}}g_{\Sigma_{b}N\bar{B}}g_{\bar{B}^{*0}B^{+}\pi^{+}}\int\frac{d^{4}k_{1}}{(2\pi)^{4}}\Phi[(p\omega_{\bar{B}^{*0}} -q\omega_{n})]\frac{i}{k_{1}^{2}-m_{B^{+}}^{2}}\frac{i(-g_{\eta\sigma}+q^{\eta}q^{\sigma}/m_{\bar{B}^{*0}}^{2})}{q^{2}-m_{\bar{B}^{*0}}^{2}}\frac{i(p\!\!\!/+m_{n})}{p^{2}-m_{n}^{2}}(i p_{2}^{\eta})\nonumber\\
	&\times\gamma^{5}+g_{\Lambda_{b}^{*} n\bar{B}^{*0}}g_{\Sigma_{b}N\bar{B}^{*}}g_{\bar{B}^{*0}B^{*+}\pi^{+}}\int \frac{d^4k_{1}}{(2\pi)^{4}}\Phi[(p\omega_{\bar{B}^{*0}}-q\omega_{n})]\epsilon_{\alpha\mu\nu\lambda}\gamma^{\theta}\frac{i(-g^{\lambda\theta}+k_{1}^{\lambda}k_{1}^{\theta}/m_{B^{*+}}^{2})}{k_{1}^{2}-m_{B^{*+}}^{2}}p_{2}^{\nu}(k_{1}^{\alpha}-q^{\alpha})\nonumber\\
    &\times \frac{i(-g^{\mu\sigma}+q^{\mu}q^{\sigma}/m_{\bar{B}^{*0}}^{2})}{q^{2}-m_{\bar{B}^{*0}}^{2}}\frac{i(p\!\!\!/+m_{n})}{p^{2}-m_{n}^{2}}\gamma^{5}\}\mu^{\sigma}(k_{0}),\\
\mathcal{M}_{b}^{3/2+}&=\frac{i}{\sqrt{2}}\mu(p_{1})\{ig_{\Lambda_{b}^{*} n\bar{B}^{*0}}g_{\Sigma_{b}N\bar{B}}g_{\bar{B}^{*0}B^{0}\pi^{0}}\int\frac{d^{4}k_{1}}{(2\pi)^{4}}\Phi[(p\omega_{\bar{B}^{*0}} -q\omega_{n})]\frac{i}{k_{1}^{2}-m_{B^{0}}^{2}}\frac{i(-g_{\eta\sigma}+q^{\eta}q^{\sigma}/m_{\bar{B}^{*0}}^{2})}{q^{2}-m_{\bar{B}^{*0}}^{2}}\frac{i(p\!\!\!/+m_{n})}{p^{2}-m_{n}^{2}}(ip_{2}^{\eta})\gamma^{5}\nonumber
\end{align}
\begin{align}	&+g_{\Lambda_{b}^{*}n\bar{B}^{*0}}g_{\Sigma_{b}N\bar{B}^{*}}g_{\bar{B}^{*0}B^{*0}\pi^{0}}\int\frac{d^4k_{1}}{(2\pi)^{4}}\Phi[(p\omega_{\bar{B}^{*(0)}}-q\omega_{n})]\times\epsilon_{\alpha\mu\nu\lambda}\gamma^{\theta}\frac{-g^{\lambda\theta}+k_{1}^{\lambda}k_{1}^{\theta}/m_{B^{*0}}^{2}}{k_{1}^{2}-m_{B^{*0}}^{2}}p_{2}^{\nu}(k_{1}^{\alpha}-q^{\alpha})\nonumber\\
	&\times\frac{i(-g^{\mu\sigma}+q^{\mu}q^{\sigma}/m_{\bar{B}^{*0}}^{2})}{q^{2}-m_{\bar{B}^{*0}}^{2}}\frac{i(p\!\!\!/+m_{n})}{p^{2}-m_{n}^{2}}\gamma^{5}\}\mu^{\sigma}(k_{0}),\\
\mathcal{M}_{c}^{3/2+}&=\frac{-1}{\sqrt{2}}i\mu(p_{1})\{ig_{\Lambda_{b}^{*} pB^{*-}}g_{\Sigma_{b}N\bar{B}}g_{B^{*-}B^{0}\pi^{-}}\int\frac{d^{4}k_{1}}{(2\pi)^{4}}\Phi[(p\omega_{B^{*-}} -q\omega_{p})]\frac{i}{k_{1}^{2}-m_{B^{0}}^{2}}\frac{i(-g_{\eta\sigma}+q^{\eta}q^{\sigma}/m_{B^{*-}}^{2})}{q^{2}-m_{B^{*-}}^{2}}\nonumber\\
	&\times\frac{i(p\!\!\!/+m_{p})}{p^{2}-m_{p}^{2}}(ip_{2}^{\eta})\gamma^{5}+g_{\Lambda_{b}^{*}pB^{*-}}g_{\Sigma_{b}N\bar{B}^{*}}g_{B^{*-}B^{*0}\pi^{-}}\int \frac{d^4k_{1}}{(2\pi)^{4}}\Phi[(p\omega_{B^{*-}}-q\omega_{p})]\epsilon_{\alpha\mu\nu\lambda}\gamma^{\theta}\frac{i(-g^{\lambda\theta}+k_{1}^{\lambda}k_{1}^{\theta}/m_{B^{*0}}^{2})}{k_{1}^{2}-m_{B^{*0}}^{2}}(k_{1}^{\alpha}-q^{\alpha})\nonumber\\
    &\times p_{2}^{\nu}\frac{i(-g^{\mu\sigma}+q^{\mu}q^{\sigma}/m_{B^{*-}}^{2})}{q^{2}-m_{B^{*-}}^{2}}\times\frac{i(p\!\!\!/+m_{p})}{p^{2}-m_{p}^{2}}\gamma^{5}\}\mu^{\sigma}(k_{0}),\\
\mathcal{M}_{d}^{3/2+}&=\frac{-1}{\sqrt{2}}i\mu(p_{1})\{ig_{\Lambda_{b}^{*} nB^{*-}}g_{\Sigma_{b}N\bar{B}}g_{B^{*-}B^{+}\pi^{0}}\int\frac{d^{4}k_{1}}{(2\pi)^{4}}\Phi[(p\omega_{B^{*-}} -q\omega_{p})]\frac{i}{k_{1}^{2}-m_{B^{+}}^{2}}\frac{i(-g_{\eta\sigma}+q^{\eta}q^{\sigma}/m_{B^{*-}}^{2})}{q^{2}-m_{B^{*-}}^{2}}\nonumber\\
	&\times\frac{i(p\!\!\!/+m_{p})}{p^{2}-m_{p}^{2}}(i p_{2}^{\eta})\gamma^{5}+g_{\Lambda_{b}^{*} pB^{*-}}g_{\Sigma_{b}N\bar{B}^{*}}g_{B^{*-}B^{*+}\pi^{0}}\int \frac{d^4k_{1}}{(2\pi)^{4}}\Phi[(p\omega_{B^{*-}} -q\omega_{p})]\epsilon_{\alpha\mu\nu\lambda}\gamma^{\theta}\frac{i(-g^{\lambda\theta}+k_{1}^{\lambda}k_{1}^{\theta}/m_{B^{*+}}^{2})}{k_{1}^{2}-m_{B^{*+}}^{2}}\nonumber\\
    &\times p_{2}^{\nu}(k_{1}^{\alpha}-q^{\alpha})\frac{i(-g^{\mu\sigma}+q^{\mu}q^{\sigma}/m_{B^{*-}}^{2})}{q^{2}-m_{B^{*-}}^{2}}\frac{i(p\!\!\!/+m_{p})}{p^{2}-m_{p}^{2}}\gamma^{5}\}\mu^{\sigma}(k_{0}),\\
\mathcal{M}_{e}^{3/2+}&=\frac{1}{\sqrt{2}}i\mu^{\theta}(p_{1})\{g_{\Lambda_{b}^{*} n\bar{B}^{*0}}g_{\Sigma_{b}^{*}N\bar{B}^{*}}g_{\bar{B}^{*0}B^{*+}\pi^{+}}\int\frac{d^4k_{1}}{(2\pi)^{4}}\epsilon_{\alpha\mu\nu\lambda}\Phi[(p\omega_{\bar{B}^{*0}}-q\omega_{n})]\frac{i(-g^{\lambda\theta}+k_{1}^{\lambda}k_{1}^{\theta}/m_{B^{*+}}^{2})}{k_{1}^{2}-m_{B^{*+}}^{2}}(k_{1}^{\alpha}-q^{\alpha})\nonumber\\
	&\times{}p_{2}^{\nu}\frac{i(-g^{\mu\sigma}+q^{\mu}q^{\sigma}/m_{\bar{B}^{*0}}^{2})}{q^{2}-m_{\bar{B}^{*0}}^{2}}\frac{i(p\!\!\!/+m_{n})}{p^{2}-m_{n}^{2}}\gamma^{5}\}\mu^{\sigma}(k_{0}),\\
\mathcal{M}_{f}^{3/2+}&=\frac{1}{\sqrt{2}}i\mu^{\theta}(p_{1})\{g_{\Lambda_{b}^{*} n\bar{B}^{*0}}g_{\Sigma_{b}^{*}N\bar{B}^{*}}g_{\bar{B}^{*0}B^{*0}\pi^{0}}\int\frac{d^4k_{1}}{(2\pi)^{4}}\epsilon_{\alpha\mu\nu\lambda}\Phi[(p\omega_{\bar{B}^{*0}}-q\omega_{n})]\frac{i(-g^{\lambda\theta}+k_{1}^{\lambda}k_{1}^{\theta}/m_{B^{*0}}^{2})}{k_{1}^{2}-m_{B^{*0}}^{2}}(k_{1}^{\alpha}-q^{\alpha})\nonumber\\
	&\times p_{2}^{\nu}\frac{i(-g^{\mu\sigma}+q^{\mu}q^{\sigma}/m_{\bar{B}^{*0}}^{2})}{q^{2}-m_{\bar{B}^{*0}}^{2}}\times\frac{i(p\!\!\!/+m_{n})}{p^{2}-m_{n}^{2}}\gamma^{5}\}\mu^{\sigma}(k_{0}),\\
\mathcal{M}_{g}^{3/2+}&=\frac{-1}{\sqrt{2}}i\mu^{\theta}(p_{1})\{g_{\Lambda_{b}^{*} pB^{*-}}g_{\Sigma_{b}^{*}N\bar{B}^{*}}g_{B^{*-}B^{*0}\pi^{-}}\int \frac{d^4k_{1}}{(2\pi)^{4}}\epsilon_{\alpha\mu\nu\lambda}\Phi[(p\omega_{B^{*-}}-q\omega_{p})]\frac{i(-g^{\lambda\theta}+k_{1}^{\lambda}k_{1}^{\theta}/m_{B^{*0}}^{2})}{k_{1}^{2}-m_{B^{*0}}^{2}}(k_{1}^{\alpha}-q^{\alpha})\nonumber\\
	&\times p_{2}^{\nu}\frac{i(-g^{\mu\sigma}+q^{\mu}q^{\sigma}/m^{2}_{B^{*-}})}{q^{2}-m^{2}_{B^{*-}}}\frac{i(p\!\!\!/+m_{p})}{p^{2}-m_{p}^{2}}\gamma^{5}\}\mu^{\sigma}(k_{0}),\\
\mathcal{M}_{h}^{3/2+}&=\frac{-1}{\sqrt{2}}i\mu^{\theta}(p_{1})\{g_{\Lambda_{b}^{*} pB^{*-}}g_{\Sigma_{b}^{*}N\bar{B}^{*}}g_{B^{*-}B^{*+}\pi^{0}}\int \frac{d^4k_{1}}{(2\pi)^{4}}\epsilon_{\alpha\mu\nu\lambda}p\Phi[(p\omega_{B^{*-}}-q\omega_{p})]\frac{i(-g^{\lambda\theta}+k_{1}^{\lambda}k_{1}^{\theta}/m_{B^{*+}}^{2})}{k_{1}^{2}-m_{B^{*+}}^{2}} p_{2}^{\nu}(k_{1}^{\alpha}-q^{\alpha})\nonumber\\
	&\times\frac{i(-g^{\mu\sigma}+q^{\mu}q^{\sigma}/m^{2}_{B^{*-}})}{q^{2}-m^{2}_{B^{*-}}}\frac{i(p\!\!\!/+m_{p})}{p^{2}-m_{p}^{2}}\gamma^{5}
\}\mu^{\sigma}(k_{0}),
\end{align}
\subsection{The decay $\Lambda_{b}(6152) $$\to$$\Sigma_{b}^{(*)}\pi$}
\begin{align}
\mathcal{M}_{a'}^{5/2+}&=\frac{-1}{\sqrt{2}}\mu(p_{1})\{ig_{\Lambda_{b}^{*} n\bar{B}^{*0}}g_{\Sigma_{b}N\bar{B}}g_{\bar{B}^{*0}B^{+}\pi^{+}}\int\frac{d^{4}k_{1}}{(2\pi)^{4}}\Phi[(p\omega_{\bar{B}^{*0}} -q\omega_{n})]\frac{i}{k_{1}^{2}-m_{B^{+}}^{2}}(i p_{2}^{\eta})(-iq^{\beta})\frac{i(-g_{\eta\sigma}+q^{\eta}q^{\sigma}/m_{\bar{B}^{*0}}^{2})}{q^{2}-m_{\bar{B}^{*0}}^{2}}\frac{i(p\!\!\!/+m_{n})}{p^{2}-m_{n}^{2}}\nonumber\\
&+g_{\Lambda_{b}^{*} n\bar{B}^{*0}}g_{\Sigma_{b}N\bar{B}^{*}}g_{\bar{B}^{*0}B^{*+}\pi^{+}}\int \frac{d^4k_{1}}{(2\pi)^{4}}\Phi[(p\omega_{\bar{B}^{*0}} -q\omega_{n})]\epsilon_{\alpha\mu\nu\lambda}p \gamma^{\theta}\frac{i(-g^{\lambda\theta}+k_{1}^{\lambda}k_{1}^{\theta}/m_{B^{*+}}^{2})}{k_{1}^{2}-m_{B^{*+}}^{2}} p_{2}^{\nu}(k_{1}^{\alpha}-q^{\alpha})(-iq^{\beta})\nonumber\\
	&\times\frac{i(-g^{\mu\sigma}+q^{\mu}q^{\sigma}/m_{\bar{B}^{*0}}^{2})}{q^{2}-m_{\bar{B}^{*0}}^{2}}\frac{i(p\!\!\!/+m_{n})}{p^{2}-m_{n}^{2}}\}\mu^{\sigma\beta}(k_{0}),
\end{align}
\begin{align}
\mathcal{M}_{b'}^{5/2+}&=\frac{-1}{\sqrt{2}}\mu(p_{1})\{ig_{\Lambda_{b}^{*} n\bar{B}^{*0}}g_{\Sigma_{b}N\bar{B}}g_{\bar{B}^{*0}B^{0}\pi^{0}}\int\frac{d^{4}k_{1}}{(2\pi)^{4}}\Phi[(p\omega_{\bar{B}^{*0}} -q\omega_{n})]\times\frac{i}{k_{1}^{2}-m_{B^{0}}^{2}} (i p_{2}^{\eta})(-iq^{\beta})\frac{i(-g_{\eta\sigma}+q^{\eta}q^{\sigma}/m_{\bar{B}^{*0}}^{2})}{q^{2}-m_{\bar{B}^{*0}}^{2}}\frac{i(p\!\!\!/+m_{n})}{p^{2}-m_{n}^{2}}\nonumber\\
&+g_{\Lambda_{b}^{*} n\bar{B}^{*0}}g_{\Sigma_{b}N\bar{B}^{*}}g_{\bar{B}^{*0}B^{*0}\pi^{0}}\int \frac{d^4k_{1}}{(2\pi)^{4}}\Phi[(p\omega_{\bar{B}^{*0}} -q\omega_{n})]\epsilon_{\alpha\mu\nu\lambda}\gamma^{\theta}\frac{i(-g^{\lambda\theta}+k_{1}^{\lambda}k_{1}^{\theta}/m_{B^{*0}}^{2})}{k_{1}^{2}-m_{B^{*0}}^{2}}\times p_{2}^{\nu}(k_{1}^{\alpha}-q^{\alpha})\nonumber\\
	&\times\frac{i(-g^{\mu\sigma}+q^{\mu}q^{\sigma}/m_{\bar{B}^{*0}}^{2})}{q^{2}-m_{\bar{B}^{*0}}^{2}}\frac{i(p\!\!\!/+m_{n})}{p^{2}-m_{n}^{2}}(-iq^{\beta})\}\mu^{\sigma\beta}(k_{0}),\\
\mathcal{M}_{c'}^{5/2+}&=\frac{1}{\sqrt{2}}\mu(p_{1})\{ig_{\Lambda_{b}^{*} pB^{*-}}g_{\Sigma_{b}N\bar{B}}g_{B^{*-}B^{0}\pi^{-}}\int\frac{d^{4}k_{1}}{(2\pi)^{4}}\Phi[(p\omega_{B^{*-}} -q\omega_{p})]\times\frac{i}{k_{1}^{2}-m_{B^{0}}^{2}}(i p_{2}^{\eta})(-iq^{\beta})\frac{i(-g_{\eta\sigma}+q^{\eta}q^{\sigma}/m_{B^{*-}}^{2})}{q^{2}-m_{B^{*-}}^{2}}\frac{i(p\!\!\!/+m_{p})}{p^{2}-m_{p}^{2}}\nonumber\\
&+g_{\Lambda_{b}^{*} pB^{*-}}g_{\Sigma_{b}N\bar{B}^{*}}g_{B^{*-}B^{*0}\pi^{-}}\int \frac{d^4k_{1}}{(2\pi)^{4}}\Phi[(p\omega_{B^{*-}} -q\omega_{p})]\epsilon_{\alpha\mu\nu\lambda}\gamma^{\theta}\frac{i(-g^{\lambda\theta}+k_{1}^{\lambda}k_{1}^{\theta}/m_{B^{*0}}^{2})}{k_{1}^{2}-m_{B^{*0}}^{2}}\times p_{2}^{\nu}(k_{1}^{\alpha}-q^{\alpha})\nonumber\\
     &\times\frac{i(-g^{\mu\sigma}+q^{\mu}q^{\sigma}m_{B^{*-}}^{2})}{q^{2}-m_{B^{*-}}^{2}}\frac{i(p\!\!\!/+m_{p})}{p^{2}-m_{p}^{2}}(-iq^{\beta})\}\mu^{\sigma\beta}(k_{0}),\\
\mathcal{M}_{d'}^{5/2+}&=\frac{1}{\sqrt{2}}\mu(p_{1})\{ig_{\Lambda_{b}^{*} pB^{*-}}g_{\Sigma_{b}N\bar{B}}g_{B^{*-}B^{+}\pi^{0}}\int\frac{d^{4}k_{1}}{(2\pi)^{4}}\Phi[(p\omega_{B^{*-}} -q\omega_{p})]\times\frac{i}{k_{1}^{2}-m_{B^{+}}^{2}}(i p_{2}^{\eta})(-iq^{\beta})\frac{i(-g_{\eta\sigma}+q^{\eta}q^{\sigma}/m_{B^{*-}}^{2})}{q^{2}-m_{B^{*-}}^{2}}\frac{i(p\!\!\!/+m_{p})}{p^{2}-m_{p}^{2}}\nonumber\\
&+g_{\Lambda_{b}^{*} pB^{*-}}g_{\Sigma_{b}N\bar{B}^{*}}g_{B^{*-}B^{*+}\pi^{0}}\int \frac{d^4k_{1}}{(2\pi)^{4}}\Phi[(p\omega_{B^{*-}} -q\omega_{p})]\epsilon_{\alpha\mu\nu\lambda}\gamma^{\theta}\frac{i(-g^{\lambda\theta}+k_{1}^{\lambda}k_{1}^{\theta}/m_{B^{*+}}^{2})}{k_{1}^{2}-m_{B^{*+}}^{2}}\times p_{2}^{\nu}(k_{1}^{\alpha}-q^{\alpha})\nonumber\\
	&\times\frac{i(-g^{\mu\sigma}+q^{\mu}q^{\sigma}/m_{B^{*-}}^{2})}{q^{2}-m_{B^{*-}}^{2}}\frac{i(p\!\!\!/+m_{p})}{p^{2}-m_{p}^{2}}(-iq^{\beta})\}\mu^{\sigma\beta}(k_{0}),\\
\mathcal{M}_{e'}^{5/2+}&=-\frac{1}{\sqrt{2}}\mu^{\theta}(p_{1})\{g_{\Lambda_{b}^{*} n\bar{B}^{*0}}g_{\Sigma_{b}^{*}N\bar{B}^{*}}g_{\bar{B}^{*0}B^{*+}\pi^{+}}\int\frac{d^4k_{1}}{(2\pi)^{4}}\epsilon_{\alpha\mu\nu\lambda}\Phi[(p\omega_{\bar{B}^{*0}}-q\omega_{n})]\frac{i(-g^{\lambda\theta}+k_{1}^{\lambda}k_{1}^{\theta}/m_{B^{*+}}^{2})}{k_{1}^{2}-m_{B^{*+}}^{2}} p_{2}^{\nu}(k_{1}^{\alpha}-q^{\alpha})\nonumber\\
	&\times\frac{i(-g^{\mu\sigma}+q^{\mu}q^{\sigma}/m_{\bar{B}^{*0}}^{2})}{q^{2}-m_{\bar{B}^{*0}}^{2}}\times\frac{i(p\!\!\!/+m_{n})}{p^{2}-m_{n}^{2}}(-iq^{\beta})\}\mu^{\sigma\beta}(k_{0}),\\
\mathcal{M}_{f'}^{5/2+}&=-\frac{1}{\sqrt{2}}\mu^{\theta}(p_{1})\{g_{\Lambda_{b}^{*} n\bar{B}^{*0}}g_{\Sigma_{b}^{*}N\bar{B}^{*}}g_{\bar{B}^{*0}B^{*0}\pi^{0}}\int\frac{d^4k_{1}}{(2\pi)^{4}}\epsilon_{\alpha\mu\nu\lambda}\Phi[(p\omega_{\bar{B}^{*0}}-q\omega_{n})]\frac{i(-g^{\lambda\theta}+k_{1}^{\lambda}k_{1}^{\theta}/m_{B^{*0}}^{2})}{k_{1}^{2}-m_{B^{*0}}^{2}} p_{2}^{\nu}(k_{1}^{\alpha}-q^{\alpha})\nonumber\\
	&\times\frac{i(-g^{\mu\sigma}+q^{\mu}q^{\sigma}/m_{\bar{B}^{*0}}^{2})}{q^{2}-m_{\bar{B}^{*0}}^{2}}\times\frac{i(p\!\!\!/+m_{n})}{p^{2}-m_{n}^{2}}(-iq^{\beta})\}\mu^{\sigma\beta}(k_{0}),\\
\mathcal{M}_{g'}^{5/2+}&=\frac{1}{\sqrt{2}}\mu^{\theta}(p_{1})\{g_{\Lambda_{b}^{*} pB^{*-}}g_{\Sigma_{b}^{*}N\bar{B}^{*}}g_{B^{*-}B^{*0}\pi^{-}}\int \frac{d^4k_{1}}{(2\pi)^{4}}\epsilon_{\alpha\mu\nu\lambda}\Phi[(p\omega_{B^{*-}}-q\omega_{p})]\frac{i(-g^{\lambda\theta}+k_{1}^{\lambda}k_{1}^{\theta}/m_{B^{*0}}^{2})}{k_{1}^{2}-m_{B^{*0}}^{2}} p_{2}^{\nu}(k_{1}^{\alpha}-q^{\alpha})\nonumber\\
	&\times\frac{i(-g^{\mu\sigma}+q^{\mu}q^{\sigma}/m^{2}_{B^{*-}})}{q^{2}-m^{2}_{B^{*-}}}\times\frac{i(p\!\!\!/+m_{p})}{p^{2}-m_{p}^{2}}(-iq^{\beta})\}\mu^{\sigma\beta}(k_{0}),\\
\mathcal{M}_{h'}^{5/2+}&=\frac{1}{\sqrt{2}}\mu^{\theta}(p_{1})\{g_{\Lambda_{b}^{*} pB^{*-}}g_{\Sigma_{b}^{*}N\bar{B}^{*}}g_{B^{*-}B^{*+}\pi^{0}}\int \frac{d^4k_{1}}{(2\pi)^{4}}\epsilon_{\alpha\mu\nu\lambda}\Phi[(p\omega_{B^{*-}}-q\omega_{p})]\frac{i(-g^{\lambda\theta}+k_{1}^{\lambda}k_{1}^{\theta}/m_{B^{*+}}^{2})}{k_{1}^{2}-m_{B^{*+}}^{2}} p_{2}^{\nu}(k_{1}^{\alpha}-q^{\alpha})\nonumber\\
	&\times\frac{i(-g^{\mu\sigma}+q^{\mu}q^{\sigma}/m^{2}_{B^{*-}})}{q^{2}-m^{2}_{B^{*-}}}\times\frac{i(p\!\!\!/+m_{p})}{p^{2}-m_{p}^{2}}(-iq^{\beta})\}\mu^{\sigma\beta}(k_{0}).
\end{align}
Once the amplitudes are determined, the corresponding partial decay widths can be obtained, which read
\begin{align}
\Gamma(\Lambda_{b}^{*}\to \pi\Sigma_b^{(*)})=\frac{1}{2J+1}\frac{1}{8\pi}\frac{|\textbf{p}_{1}|}{m^2_{\Lambda_{b}^{*}}}\overline{|{\cal{M}}|^2}\label{eq31},
\end{align}	
where the $J$ is the total angular momentum of the $\Lambda_{b}^{*}$ ,  $|\textbf{p}_{1}|$ is the three-momenta of the decay products in the center of mass frame, the overline indicates the sum over the polarization vectors of the final hadrons.  ${\cal{M}}$ is the total amplitudes, which is the sum of the individual amplitude.
	
\section{Results and discussions}\label{Sec: results}
Through the study of $\Lambda_b^{*}\to\pi\Sigma_b^{(*)}$ two-body strong decay widths, our focus in this work is on examining whether the observed $\Lambda_b(6146)$ and $\Lambda_b(6152)$
states~\cite{LHCb:2019soc} can be interpreted as predominantly $\bar{B}^{*}N$ molecules, as first proposed by Ref.~\cite{Jian:2022rln}. In order to obtain the decay widths shown in
Figs.\ref{a1} and \ref{a2}, the coupling constant $g_{\Lambda^{}_b\bar{B}^{*}N}$ should be computed first.  According to the compositeness condition introduced in Eq.~\ref{eq8}, the detailed
expression for the coupling constants as a function of the parameter $\Lambda$ is computed, which can be found in Eqs.~\ref{eq11} and \ref{eq10}.
\begin{figure}[h!]
    \centering
	\includegraphics[clip,scale=0.55]{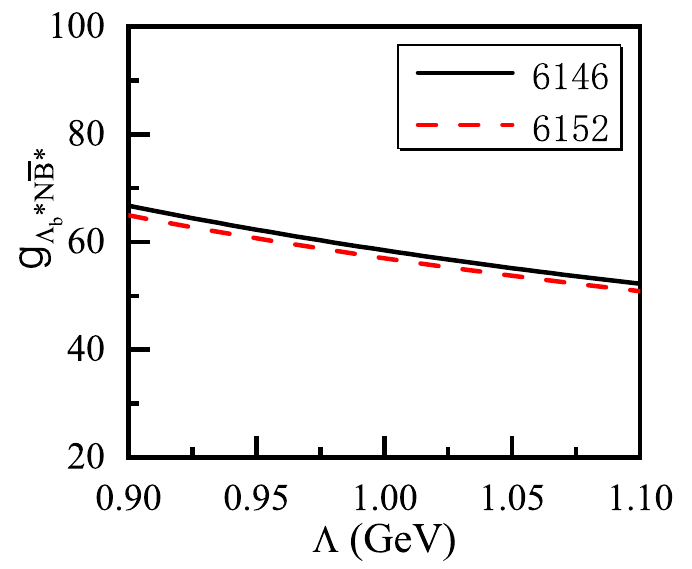}
	\caption{The coupling constant of $g_{\Lambda_{b}^{*}N\bar{B}^{*}}$ as a function of the parameter $\Lambda$.}
	\label{a88}
\end{figure}

With a cutoff value of $\Lambda=0.9-1.1$ GeV, the corresponding coupling constants are shown in Fig.\ref{a88}. It can be observed that the coupling constants $g_{\Lambda^{}_b\bar{B}^{*}N}$
decrease as the cutoff increases, indicating a strong dependence of the coupling constants on $\Lambda$. Here, the $\Lambda_b(6146)$ and $\Lambda_b(6152)$ are identified as $P$-wave and
$F$-wave $\bar{B}^{*}N$ molecular states, respectively.  This might be the reason why the coupling constants depends on the parameter $\Lambda$, which characterizes the distribution of the
components within the molecule.  It means that the $\Lambda$ is related to the relative motion of the two molecular components.  Indeed, for $S$-wave molecules, the obtained coupling constant
is only weakly dependent on $\Lambda$~\cite{Huang:2018wgr,Zhu:2021exs,Xiao:2016mho,Dong:2016dkh,Dong:2008gb,Dong:2017rmg}.

To provide a better understanding of how the coupling constants vary with the parameter $\Lambda$, we specify the ranges of their variation. When considering $\Lambda_b(6146)$ and $\Lambda_b(6152)$
as $P$-wave and $F$-wave $\bar{B}^{*}N$ molecular states, their coupling constants range from 52.22 to 66.73 and from 50.83 to 64.93, respectively. Typically, a value of $\Lambda=1.0$ GeV is used,
as seen in Refs.\cite{Huang:2018wgr,Zhu:2021exs,Xiao:2016mho,Dong:2016dkh,Dong:2008gb,Dong:2017rmg}.  Therefore, in this work, we adopt $\Lambda=1.0$ GeV, resulting in corresponding coupling constants
of $g_{\Lambda_b(6146)\bar{B}^{*}N}=58.44$ and $g_{\Lambda_b(6152)\bar{B}^{*}N}=56.97$.  The slight difference in the coupling constants $g_{\Lambda_b(6146)\bar{B}^{}N}$ and $g_{\Lambda_b(6152)\bar{B}^{}N}$
can be easily understood due to the small disparity in the masses of these two states, approximately 8 MeV. Additionally, we observed that despite the higher mass of $\Lambda_b(6152)$ compared to
$\Lambda_b(6146)$, its coupling to the molecular component $\bar{B}^{*}N$ is weaker. One possible explanation for this is that the mass of $\Lambda_b(6152)$ is closer to the $\bar{B}^{*}N$ threshold
compared to $\Lambda_b(6146)$, making $\Lambda_b(6152)$ a relatively looser $\bar{B}^{*}N$ molecular state compared to $\Lambda_b(6146)$.

\begin{figure}[h!]
    \centering
	\includegraphics[clip,scale=0.55]{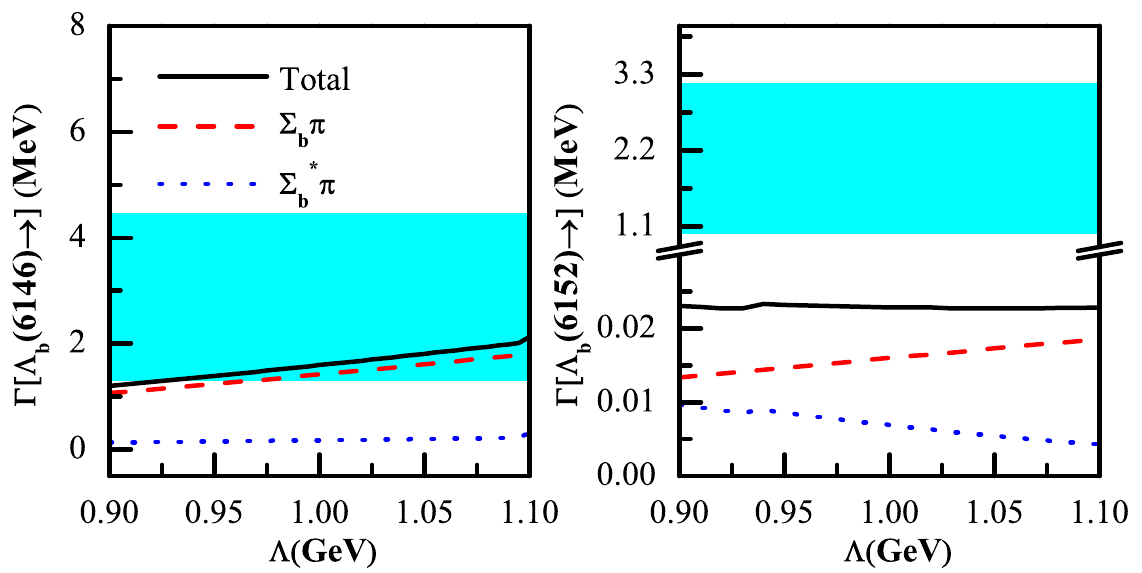}
	\caption{Partial decay widths of the $\Lambda_{b}(6146/6152)\to\Sigma_{b}\pi$ (red dashed line), $\Lambda_{b}(6146/6152)\to\Sigma^{*}_{b}\pi$ (blue dotted line), and total decay widthes (black sold line).  The cyan bands
denote the experimental total width~\cite{LHCb:2019soc}.}
	\label{a5}
\end{figure}

Once the coupling constants $g_{\Lambda_b(6146)\bar{B}^{*}N}$ and $g_{\Lambda_b(6152)\bar{B}^{*}N}$ are obtained, the decay widthes for the $\Lambda_b(6146)\to\Sigma_b^{(*)}\pi$ and
$\Lambda_b(6152)\to\Sigma_b^{(*)}\pi$ can be easily calculated.  In Fig.~\ref{a5}, we show the total decay widthes of $\Lambda_b(6146)\to\Sigma_b^{(*)}\pi$ and $\Lambda_b(6152)\to\Sigma_b^{(*)}\pi$
as a function of the cutoff parameter $\Lambda$.  Furthermore, for a clear comparison, the experimental data~\cite{LHCb:2019soc}, indicated by cyan bands, are also plotted in Fig.~\ref{a5}.
In the present calculation, we vary $\Lambda$ from 0.9 to 1.1 GeV.   In this cutoff range, the total decay width increases for the $\Lambda_b(6146)\to\Sigma_b^{(*)}\pi$ decay, while decreasing
slowly for the $\Lambda_b(6152)\to\Sigma_b^{(*)}\pi$ decay.  The results also indicate that the transitions $\Lambda_{b}(6146/6152)\to\Sigma_{b}\pi$ are the main decay channel, which almost saturates
the total width of  $\Lambda_b(6146/6152)$.  However, the transition $\Lambda_{b}(6146/6152)\to\Sigma^{*}_{b}\pi$  gives minor contributions, with transition strengths on the order of about 0.1 MeV
and 0.001 MeV, respectively.

Now we present the detailed numerical results in Tab.~\ref{tab-2} and the corresponding conclusions.  For the $\Lambda_b(6146)\to\Sigma_b^{(*)}\pi$ decay, the predicted total decay width increases from
1.20 to 2.23 MeV (see Tab.~\ref{tab-2}), aligning with the experimental total width.  This suggests that the decay width of the observed $\Lambda_b(6146)$ can be well reproduced in the $\bar{B}^{*}N$
molecular picture, supporting its characterization as a $p$-wave $\bar{B}^{*}N$ molecular state. Additionally, the results also tell us that the transition $\Lambda_{b}(6146)\to\Sigma_{b}\pi$  serves as the
predominant decay channel, nearly fully accounting for the total width of $\Lambda_b(6146)$, consistent with experimental observations that emphasize the $\Lambda_b(6146)$ decay primarily into the $\pi\Sigma_b$
channel (see Fig.~3 of Ref.~\cite{LHCb:2019soc}). This also provides direct evidence that the observed $\Lambda_b(6146)$ is a $p$-wave $\bar{B}^{*}N$ molecular state.
\begin{table*}[http!]
\centering
\caption{ Partial decay widths of $\Lambda_b^{*}\to\Sigma_b^{(*)}\pi$ and the total decay width with $\Lambda=0.9-1.1$ GeV and its center value, in comparison with the results of the quark model~\cite{Wang:2019uaj,Azizi:2020tgh,Chen:2019ywy,Liang:2019aag} and total width obtained from the LHCb experiments~\cite{LHCb:2019soc}. The symbol $\checkmark$ represents this channel play a major contribution. All the widths are in units of MeV.  }\label{tab-2}
\setlength{\tabcolsep}{2mm}{
\begin{tabular}{cccccccccc}
\hline\hline
 Model                                 &~~~~~ $\Lambda=1.0$      &~~~~~ $\Lambda=0.9-1.1$      &~~~~~ Exp.~\cite{LHCb:2019soc}   &~~~~~ Ref.~\cite{Liang:2019aag}  &~~~~~ Ref.~\cite{Chen:2019ywy}&~~~~~ Ref.~\cite{Azizi:2020tgh} &~~~~~ Ref.~\cite{Wang:2019uaj}~~~~~ \\ \hline
  $\Lambda_b(6146)\to{}\pi\Sigma_b$    &~~~~~ $1.42$                &~~~~~ $1.07-1.81$               &~~~~~ $\checkmark$               &~~~~~ $5.31$                     &~~~~~3.25                     &~~~~~$2.3\pm{}0.4$         &~~~~~ 4.41\\
  $~~~~~~~~~~~~~~~\to{}\pi\Sigma^{*}_b$&~~~~~  $0.17$               &~~~~~ $0.13-0.42$               &~~~~~ $     $                    &~~~~~ $0.87$                     &~~~~~0.93                     &~~~~~$0.7\pm0.1$           &~~~~~ 1.26 \\
  $~~~~~~~~~~~~~~~Total$               &~~~~~  $1.59$               &~~~~~ $1.20-2.23$               &~~~~~ $2.9\pm1.3\pm0.3$          &~~~~~ $6.17$                     &~~~~~4.18                     &~~~~~$3.0\pm0.4$           &~~~~~ 5.67\\ \hline
  $\Lambda_b(6152)\to{}\pi\Sigma_b$    &~~~~~  $0.0160$             &~~~~~ $0.0134-0.0186$           &~~~~~ $     $                    &~~~~~ $0.03$                     &~~~~~0.22                     &~~~~~                      &~~~~~0.73\\
  $~~~~~~~~~~~~~~~\to{}\pi\Sigma^{*}_b$&~~~~~  $0.0069$             &~~~~~ $0.0043-0.0097$           &~~~~~ $\checkmark$               &~~~~~ $5.41$                     &~~~~~4.17                     &~~~~~                      &~~~~~4.60\\
  $~~~~~~~~~~~~~~~Total$               &~~~~~  $0.0229$             &~~~~~ $0.0228-0.0233$           &~~~~~ $2.1\pm0.8\pm0.3$          &~~~~~ $5.44$                     &~~~~~4.39                     &~~~~~                      &~~~~~5.33\\
\hline \hline
\end{tabular}}
\end{table*}

The results also indicate that considering $\Lambda_b(6152)$ as an $F$-wave $\bar{B}^{*}N$ molecule is not viable (see Fig.~\ref{a5}). This is evident from the calculated decay width,
$\Gamma_{\Lambda_b(6152)}=0.0288-0.0233$  MeV (see Tab.~\ref{tab-2}), which is significantly smaller than the experimental total width, $\Gamma_{\Lambda_{b}(6152)}=2.1\pm0.8\pm0.3$ MeV~\cite{LHCb:2019soc}.
Meanwhile, the main decay channel we computed is $\pi\Sigma_b$, which contradicts the experimental observation that the primary decay contribution comes from the $\pi\Sigma_b^{*}$ channel.  These results may
serve as evidence that $\Lambda_b(6152)$ is a conventional baryon containing a $b$ quark.

\begin{figure}[h!]
    \centering
	\includegraphics[clip,scale=0.45]{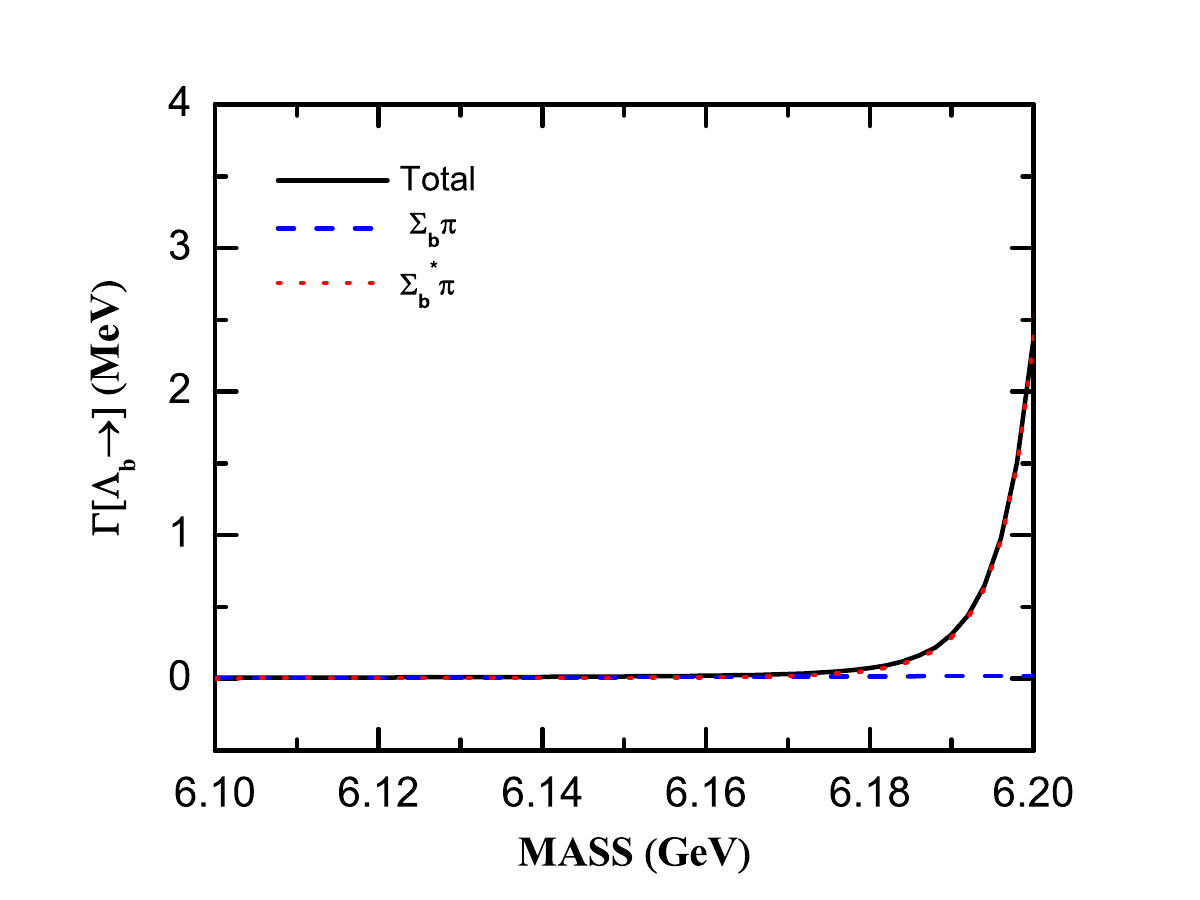}
	\caption{The decay widths of $\Lambda_b^{*}$ with $J^p=5/2^{+}$ are plotted as functions of the masses of the $\bar{B}^{*}N$ molecules at $\Lambda=1.0$ GeV.}
	\label{a98}
\end{figure}
Our study found that the observed $\Lambda_b(6146)$ is a $p$-wave $\bar{B}^{*}N$ molecular state, while disfavors the assignment of $\Lambda_b(6152)$ as a $F$-wave $\bar{B}^{*}N$  molecule.  This suggests
that the internal structures of $\Lambda_b(6146)$ and $\Lambda_b(6152)$ are very different, making it unlikely that they form a pair that satisfies heavy quark spin symmetry~\cite{Isgur:1991wq}.
To identify the heavy quark spin symmetry partners for $\Lambda_b(6146)$, we computed the decay widthes of the $\bar{B}^{*}N$ molecule with $J^p=5/2^{+}$ as a function of its masses at $\Lambda=1.0$ GeV$^{[1]}$
\footnotetext[1]{This is because the decay widths of the two particles, which satisfy heavy quark spin symmetry, are almost identical~\cite{Isgur:1991wq}.}.
The results are plotted in Fig.~\ref{a98}, and we can find that within the range of $\bar{B}^{*}N$ molecule masses from 6195 to 6200 MeV, its decay width can vary from 1.0 to 2.5 MeV. Furthermore, within
this mass range, its primary decay width is into $\pi\Sigma_b^{*}$ channel.  It seems to suggest that baryons with a $\bar{B}^*N$ molecular structure, which are the heavy quark spin symmetry partners of $\Lambda_b(6146)$,
can be searched for in the mass range of 6195-6200 MeV.  This implies that not all pairs of hadrons containing a heavy quark (c or b), with a mass difference of only a few MeV and almost equal decay widths,
necessarily form heavy quark spin symmetry doublet states.

According to the heavy quark flavor symmetry, the observed $\Lambda_b(6146)$ can be considered a $p$-wave $\bar{B}^{*}N$ molecular state, suggesting the existence of a $D^{*}N$ molecule with $J^p=3/2^{+}$.
This $D^{*}N$ molecule could correspond to the experimentally observed $\Lambda_c(2860)^{+}$. If $\Lambda_c(2880)^{+}$ is indeed the heavy quark flavor symmetry partner of $\Lambda_b(6152)$, it would
possess a conventional three-quark structure.  Consequently, we suggest seeking a $D^{*}N$ molecule with a spin-parity of $J^p=5/2^{+}$, which is the heavy-quark spin partner state of $\Lambda_c(2860)^{+}$.

It was worth noting that although our results could support the idea that $\Lambda_b(6146)$ is a $\bar{B}^{*}N$ molecule, the calculated total width, particularly with $\Lambda=1.0$, yields a width of $\Gamma_{\Lambda_b(6146)}=1.59$ MeV (see tab.~\ref{tab-2}), which is smaller than the experimental central value of $\Gamma_{\Lambda_b(6146)}=2.9$ MeV.  This tells us that $\Lambda_b(6146)$ might not
solely consist of a $P$-wave $\bar{B}^{*}N$ molecule; it could also contain additional three-quark components, a structure permissible within quantum chromodynamics.  The rough estimate indicates that
the $\bar{B}^{*}N$ molecular component of $\Lambda_b(6146)$ accounts for approximately 54.8\% of its total component (calculated as 1.59/2.9).

Evidence supporting the conclusion  that $\Lambda_b(6146)$
displays a mixed structure can be also found in many prior studies, including those cited in Refs.~\cite{Wang:2019uaj,Liang:2019aag}.  Although considering $\Lambda_b(6146)$ as a $1D$-wave resonance with
$J^p=3/2^{+}$ suggests that the dominant decay mode of this state is $\Lambda_{b}(6146)\to\Sigma_{b}\pi$, the total decay width obtained via the quark model in these two works is slightly larger than the
experimental data within the reported errors (see the fifth and eighth columns of Tab.~\ref{tab-2}), with an excess over the experimental maximum data of about 37.1\% [(6.17-4.5)/4.5] and 26.0\% [(5.67-4.5)/4.5],
respectively.  These results implies that $\Lambda_b(6146)$ is not a pure three-quark state, but contains additional components, possibly including the $\bar{B}^{*}N$ molecule proposed by us in this work
and also in Ref.~\cite{Jian:2022rln}.  Note that the conclusion of a pure three-quark state structure for $\Lambda_b(6146)$ can be derived from Refs.~\cite{Chen:2019ywy,Azizi:2020tgh}, as the main decay
channel and the total decay width obtained in these two works are both consistent with the experimental observations.

The same conclusion regarding mixed structure can be also inferred for $\Lambda_b(6146)$.  The total decay width obtained based on the quark model is 5.33 MeV~\cite{Wang:2019uaj}, 4.39 MeV~\cite{Chen:2019ywy},
or 5.44 MeV~\cite{Liang:2019aag}, which are not within the experimental measurement range and are slightly larger (see tab.~\ref{tab-2}).  Combining these results, we can infer that $\Lambda_b(6152)$ may
primarily consist of three quark component and contain a small fraction of molecular state components.

\section{Summary}\label{sec:summary}
In the present work, we provide the detailed derivation and formalism for calculating the decay width of the $\Lambda_b(6146) \to \Sigma^{(*)}_b \pi$ and $\Lambda_b(6152) \to \Sigma^{(*)}_b \pi$ processes. Our theoretical starting point is to consider $\Lambda_b(6146)$ and $\Lambda_b(6152)$ as molecules primarily composed of the $\bar{B}^{*}N$ component, a result first given in Ref.~\cite{Jian:2022rln}. Based on this molecular assumption, the decay process can be summarized as follows: first, $\Lambda_b(6146)$ and $\Lambda_b(6152)$ couple with their molecular component $\bar{B}^{*}N$, and then decay to the $\pi \Sigma_b$ and $\pi \Sigma_b^{*}$ final states through
the $\bar{B}^{*}N$ molecular component, which includes a hadronic loop. The corresponding Feynman diagrams for the processes are shown in Fig. \ref{a1} and Fig. \ref{a2}.

Our calculations show that within the parameter range of $\Lambda = 0.9-1.0$ GeV, the total decay width of $\Lambda_b(6146)$ is 1.20-2.23 MeV, with $\Lambda_b(6146) \to \pi \Sigma_b$ being the primary decay channel. This aligns with experimental measurements~\cite{LHCb:2019soc}, supporting the identification of $\Lambda_b(6146)$ as an $\bar{B}^{*}N$ molecular state.  However, our calculations do not support $\Lambda_b(6152)$ as an $\bar{B}^{*}N$ molecular state.  The theoretically obtained total decay width of $\Lambda_b(6152)$ is $0.0228-0.0233$ MeV, which is significantly lower than the experimental value.  Additionally, its decay is primarily through the $\Sigma_b \pi$ channel, which contrasts with experimental findings that indicate the main contribution comes from the $\Sigma_b^{*} \pi$ channel~\cite{LHCb:2019soc}.

Under the guidance of heavy quark symmetry, $\Lambda_b(6146)$ has the properties of a $P$-wave $N\bar{B}^*$ molecule, suggesting the existence of a $D^{*}N$ molecule with spin $J^P = 3/2^+$, which resonates with the experimentally observed $\Lambda_c(2860)^+$. If $\Lambda_c(2880)^+$ is confirmed to be the heavy quark flavor symmetry partner of $\Lambda_b(6152)$, it is expected to have a typical three-quark structure. In the mass range of 6195$\sim$6200 MeV, we believe that there may be an $\bar{B}^{*}N$ molecular state with $J^P=5/2^{+}$ as the heavy quark spin symmetry partner of $\Lambda_b(6146)$. Therefore, we suggest conducting experiments to search for this $b$-quark-containing baryon and its heavy quark flavor symmetry partner.   This also tell us that not all pairs of hadrons containing a heavy quark (c or b), with a mass difference of only a few MeV and almost equal decay widths,
necessarily form heavy quark spin symmetry doublet states.

It should be noted that $\Lambda_b(6146)$ and $\Lambda_b(6152)$ can also be considered as conventional three-quark states~\cite{Wang:2019uaj,Azizi:2020tgh,Chen:2019ywy,Liang:2019aag,Yu:2021zvl,Kakadiya:2021jtv}. Combining these results, we infer that these two states and their heavy quark  symmetry partner states are mixed states containing both three-quark structures and molecular components. To confirm such mixed internal structure for these baryons, theoretical investigations on other decay modes and further experimental information on their partial decay widths will be very helpful.

\section*{Acknowledgments}
This work was supported by the National Natural Science Foundation
of China under Grant No.12005177.  Yin. Huang also acknowledges the support from the Fundamental Re
search Funds for the Central Universities under Grant No.
2682026TPY011.

%\end{CJK*}
%
\end{document}